\documentclass[aps,pra,noshowpacs,twocolumn,preprintnumbers,superscriptaddress,floatfix]{revtex4}
\usepackage{color}
\usepackage{amsmath}
\usepackage{graphicx}% Include figure files
\usepackage{dcolumn}% Align table columns on decimal point
\usepackage{bm}% bold math
\usepackage{hyperref,graphicx}
\usepackage{textcomp}
\usepackage{multirow}
\usepackage{dcolumn}% Align table columns on decimal point

\begin{document}

\title{Experimental Quantum-enhanced Cryptographic Remote Control}

\author{Xiao-Ling Pang}
\affiliation{State Key Laboratory of Advanced Optical Communication Systems and Networks, School of Physics and Astronomy, Shanghai Jiao Tong University, Shanghai 200240, China}
\affiliation{Synergetic Innovation Center of Quantum Information and Quantum Physics, University of Science and Technology of China, Hefei, Anhui 230026, China}
\author{Lu-Feng Qiao}
\affiliation{State Key Laboratory of Advanced Optical Communication Systems and Networks, School of Physics and Astronomy, Shanghai Jiao Tong University, Shanghai 200240, China}
\affiliation{Synergetic Innovation Center of Quantum Information and Quantum Physics, University of Science and Technology of China, Hefei, Anhui 230026, China}
\author{Ke Sun}
\affiliation{State Key Laboratory of Advanced Optical Communication Systems and Networks, School of Physics and Astronomy, Shanghai Jiao Tong University, Shanghai 200240, China}
\affiliation{Zhiyuan Innovative Research Center, Shanghai Jiao Tong University, Shanghai 200240, China}
\author{Yu Liu}
\affiliation{State Key Laboratory of Advanced Optical Communication Systems and Networks, School of Physics and Astronomy, Shanghai Jiao Tong University, Shanghai 200240, China}
\affiliation{Zhiyuan Innovative Research Center, Shanghai Jiao Tong University, Shanghai 200240, China}
\author{Ai-Lin Yang}
\affiliation{State Key Laboratory of Advanced Optical Communication Systems and Networks, School of Physics and Astronomy, Shanghai Jiao Tong University, Shanghai 200240, China}
\affiliation{Synergetic Innovation Center of Quantum Information and Quantum Physics, University of Science and Technology of China, Hefei, Anhui 230026, China}
\author{Xian-Min Jin}
\email{xianmin.jin@sjtu.edu.cn}
\affiliation{State Key Laboratory of Advanced Optical Communication Systems and Networks, School of Physics and Astronomy, Shanghai Jiao Tong University, Shanghai 200240, China}
\affiliation{Synergetic Innovation Center of Quantum Information and Quantum Physics, University of Science and Technology of China, Hefei, Anhui 230026, China}

\begin{abstract}
The Internet of Things (IoT), as a cutting-edge integrated cross-technology, promises to informationize people's daily lives, while being threatened by continuous challenges of eavesdropping and tampering. The emerging quantum cryptography, harnessing the random nature of quantum mechanics, may also enable unconditionally secure control network, beyond the applications in secure communications. Here, we present a quantum-enhanced cryptographic remote control scheme that combines quantum randomness and one-time pad algorithm for delivering commands remotely. We experimentally demonstrate this on an unmanned aircraft vehicle (UAV) control system. We precharge quantum random number (QRN) into controller and controlee before launching UAV, instead of distributing QRN like standard quantum communication during flight. We statistically verify the randomness of both quantum keys and the converted ciphertexts to check the security capability. All commands in the air are found to be completely chaotic after encryption, and only matched keys on UAV can decipher those commands precisely. In addition, the controlee does not response to the commands that are not or incorrectly encrypted, showing the immunity against interference and decoy. Our work adds true randomness and quantum enhancement into the realm of secure control algorithm in a straightforward and practical fashion, providing a promoted solution for the security of artificial intelligence and IoT.
\end{abstract}

\maketitle 

%\textbf{The Internet of Things (IoT), as a cutting-edge integrated cross-technology, promises to informationize people's daily lives, while being threatened by continuous challenges of eavesdropping and tampering. The emerging quantum cryptography, harnessing the random nature of quantum mechanics, may also enable unconditionally secure control network, beyond the applications in secure communications. Here, we present a quantum-enhanced cryptographic remote control scheme that combines quantum randomness and one-time pad algorithm for delivering commands remotely. We experimentally demonstrate this on an unmanned aircraft vehicle (UAV) control system. We precharge quantum random number (QRN) into controller and controlee before launching UAV, instead of distributing QRN like standard quantum communication during flight. We statistically verify the randomness of both quantum keys and the converted ciphertexts to check the security capability. All commands in the air are found to be completely chaotic after encryption, and only matched keys on UAV can decipher those commands precisely. In addition, the controlee does not response to the commands that are not or incorrectly encrypted, showing the immunity against interference and decoy. Our work adds true randomness and quantum enhancement into the realm of secure control algorithm in a straightforward and practical fashion, providing a promoted solution for the security of artificial intelligence and IoT.}

\noindent With the rapid development of artificial intelligence and IoT, greater demands are being placed on the security by growing hacking incidents. To implement cryptographic remote control, two general types of key-based algorithms, public-key and symmetric, are being widely investigated. Public-key algorithms use two different keys for encryption and decryption, and are often based on computational complexity; while they are imperfect in the real world for being slow, and vulnerable to chosen-plaintext attacks \cite{Schneier_1996}. Conventional symmetric algorithms require that communication parties share matched and secret keys in advance; while the security of such algorithms rely on the shared keys. One-time pad \cite{Gingerich_Isis_1996}, as a powerful symmetric algorithm, has been proved by Claude Shannon to be impossible to crack \cite{Shannon_Bell_1949}, as long as crucial problems of generating and sharing real random sequences are settled.

Randomness is a fundamental resource with significant applications in cryptography and numerical simulation. Real random sequences, however, are hard to generate mathematically \cite{Knuth_1981}, but have to rely on unpredictable physical processes \cite{Stefanov_JMO_2000,Jennewein_RSI_2000,Uchida_nphoton_2008,Dynes_APL_2008,Fiorentino_PRA_2006}. Although different mechanics, such as chaotic effects \cite{Stojanovski_IEEE_2001_partI,Stojanovski_IEEE_2001_partII}, thermal noise \cite{Petrie_Computers_2004}, biometric parameters \cite{Szczepanski_IEEE_2000} and free-running oscillators \cite{Kohlbrenner_2004} are employed in the generation of physical random number, they are faced with some problems like hard to detect failure \cite{Herrero_RMP_2017}. The inherent uncertainty of quantum mechanics makes quantum systems an excellent stochastic source, with the fact that a single photon incident on a 50:50 beam splitter be transmitted or reflected is intrinsically random. More importantly, the randomness is precisely balanced and immune to environmental perturbations.

\begin{figure*}
	\centering
	\includegraphics[width=1.6\columnwidth]{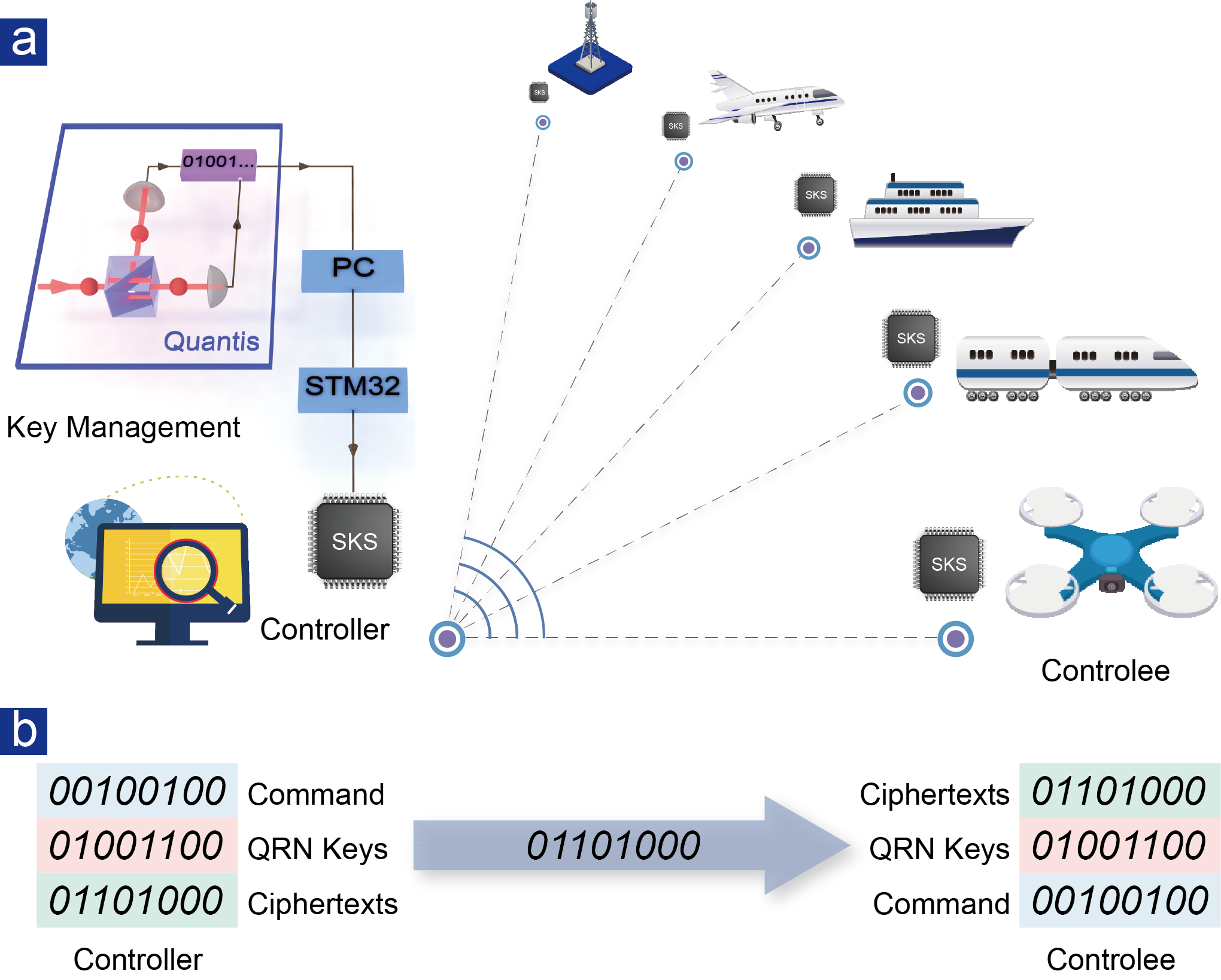}\\
	\caption{\textbf{Quantum-enhanced cryptographic remote control.} \textbf{(a)} A schematic diagram of quantum-enhanced cryptographic remote control system and the potential applications. The quantum random number generator utilized here is the ``Quantis USB" of ID Quantique company. Scenarios applicable to our direct-charging scheme: remote control of UAV, high-speed rail scheduling, vessel movement, airport dispatch and smart grid control. \textbf{(b)} A specific example that combines QRN keys and one-time pad algorithm for delivering commands remotely}
	\label{f1}
\end{figure*}

Sharing randomness is another crucial problem to be settled for realizing symmetric algorithms. One best-known scheme is quantum key distribution (QKD), which is quite mature so far for applications, with enormous progresses \cite{Bennett_IEEE_1984,Ekert_PRL_1991,Lo_nphoton_2014,JiLing_OE_2016} and is even ready for constructing secure networks \cite{Qiu_Nature_2014,Elliott_nCommun_2006,Fujiwara_OE_2011,Frohlich_PRA_2013,Tang_PRX_2016}. Nevertheless, many situations of IoT control are not compatible with QKD schemes. For example, sensor networks require low cost, low power and miniature devices, which is hard to be met by QKD systems \cite{Gubbi_FGCY_2012}, especially for large-scale and distributed sensor networks. 

Interestingly, in many situations, real-time sharing of randomness is not really necessary in practice. For instance, UAVs or satellites are essentially well identified before being launched, and are well isolated with other parties during their missions. For all these situations, we could precharge quantum keys into controllers and controlees, and implement cryptographic remote control with quantum enhancement in a straightforward and practical way. In this work, we present this quantum-enhanced cryptographic remote control scheme that combines quantum randomness and one-time pad algorithm for delivering commands remotely, and experimentally demonstrate this on a UAV control system.

The cryptographic system is composed of three subsystems: key management unit, controller unit and controlee unit, as is shown in Figure 1(a). The randomness derived from quantum nature of single photons is charged into IoT devices through secure key storage (SKS) chips to perform one-time pad encryption and decryption. The connection of flexible small-scale SKS chips to QRN generator is realized by a home-built key management unit. SKS chips are planted into controllers and controlees before they are detached. Commands are encrypted by one-time pad algorithm with a certain section of keys, which can only be decrypted correctly by the controlee with corresponding keys. On the controller side, with QRN keys, a bitwise exclusive OR is performed on commands before sent; and conversely, with identical keys, commands can be deciphered and executed on the controlee side, as is shown in Figure 1(b).

\begin{figure*}
	%\centering
	\includegraphics[width=1.65\columnwidth]{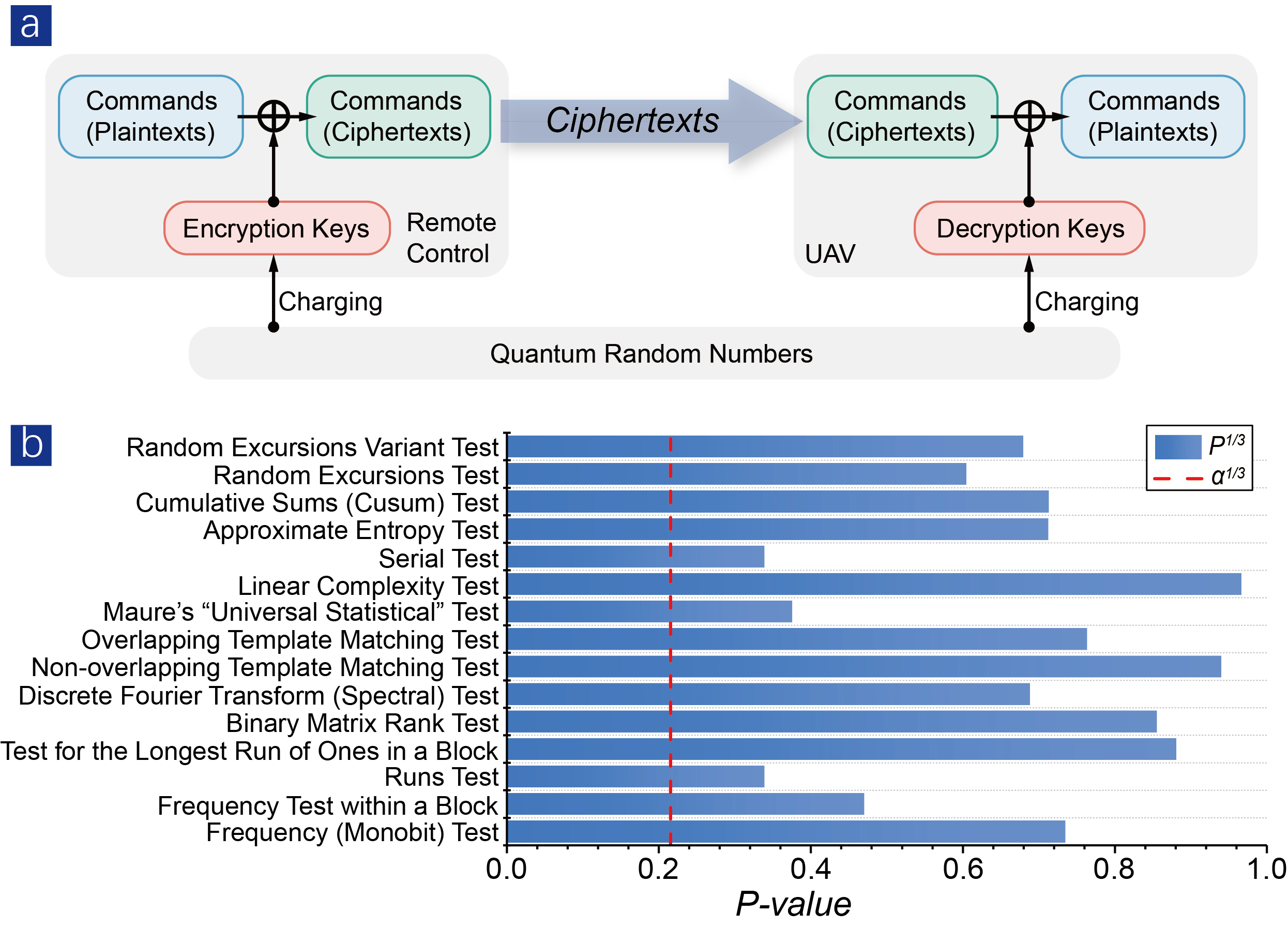}\\
	\caption{\textbf{Signal flow diagram and NIST tests of quantum randomness}. \textbf{(a)} Signal flow diagram of the quantum-enhanced cryptographic control system. \textbf{(b)} NIST statistical randomness tests performance of quantum keys. The experimental results are obtained from 1,638,400 bits samples with a significance level of $\alpha=0.01$. In the histogram, the value of each test that exceed the red dashed line turns out a successful pass.}
	\label{f2}
\end{figure*}

\begin{figure*}
	\centering
	\includegraphics[width=1.35\columnwidth]{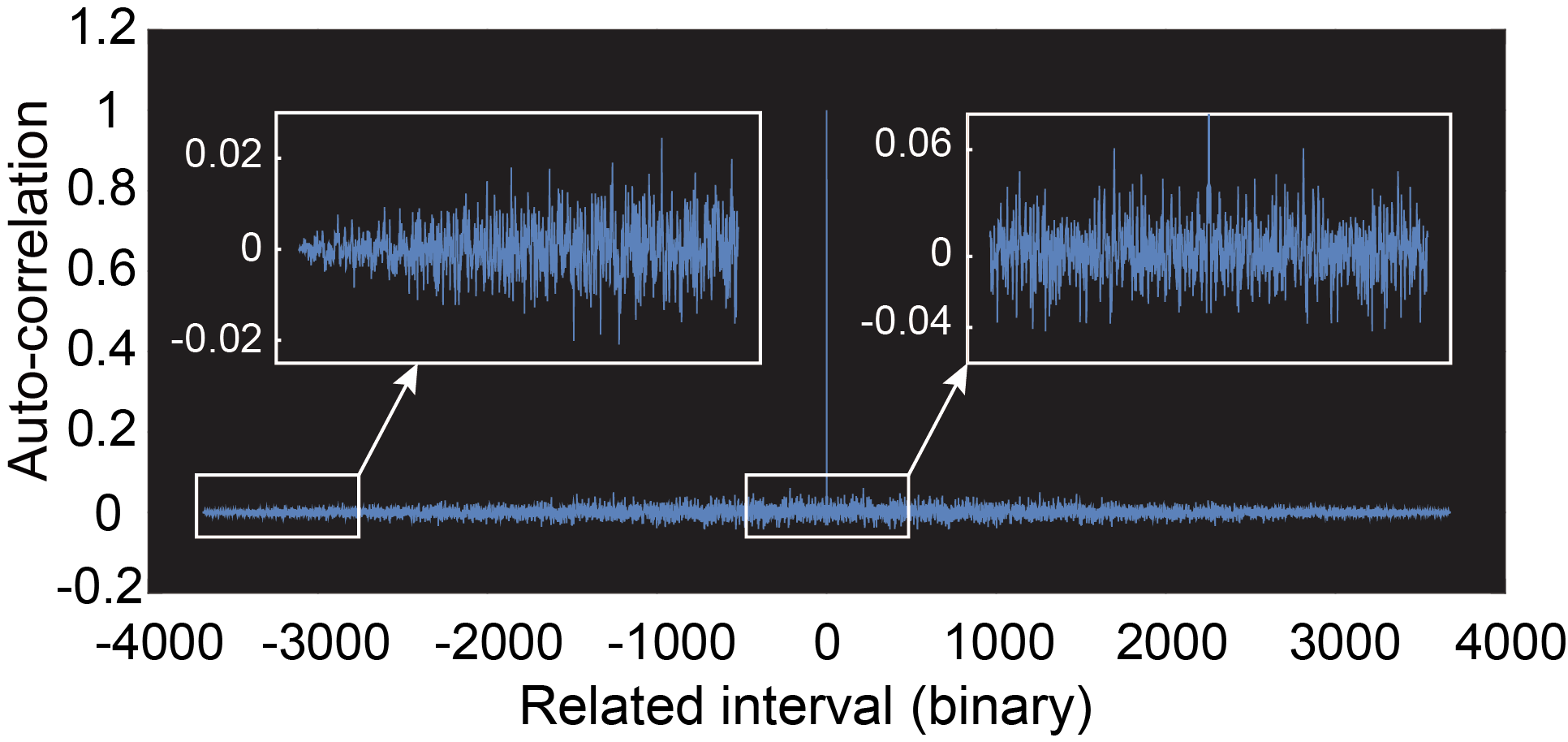}\\
	\caption{\textbf{The auto-correlation function measured among long commands ciphertexts.} The auto-correlation function of ideal random sequence is close to delta function. The sharp auto-correlation peak in the center indicates that the encrypted binary sequence has excellent independence on each part. The insets show auto-correlation details in the near- and far-field regime.}
	\label{f3}
\end{figure*}

The data transmission diagram of our UAV control system is illustrated in Figure 2(a). Encryption keys are generated by Quantis, which is a reliable QRN generator, employing a quantum process as the source of randomness, and producing random sequences at a bit rate of 4 Mb/s. To be specific, a photon incident on a semi-transparent mirror will be reflected with half the probability, leading to a ``0"; or transmitted with half the probability, leading to an ``1". A microcontroller is dedicated for charging or updating quantum keys into SKS chips.

\begin{table}[!b]
	\centering
	\caption{\textbf{Balance and runs properties.} $P\text{-}values$ for uniformity check, and proportions for examination of the sequences that pass a certain statistical test (Success Rate). 20 pieces of commands are tested.}
	
	\label{tab1}
	\begin{tabular}{p{2.1cm}<{\centering} p{2.0cm}<{\centering} p{2.0cm}<{\centering}  p{2.0cm}<{\centering}}
		\hline\noalign{\vskip 0.6mm}
		\hline\noalign{\vskip 0.14cm}
		
		Test Index &  $P\text{-}value$ & Proportion & Result  \\[0.07cm]
		
		\hline\noalign{\vskip 0.12cm}
		
		Frequency	& 0.4861  &	1	&	SUCCESS           \\[0.06cm]
		Runs		& 0.4719  &	1	&	SUCCESS           \\[0.06cm]
				
		\hline\noalign{\vskip 0.6mm}
		\hline\noalign{\vskip 0.6mm}
	\end{tabular}
	\par
\end{table}

\begin{figure*}
	%\centering
	\includegraphics[width=1.65\columnwidth]{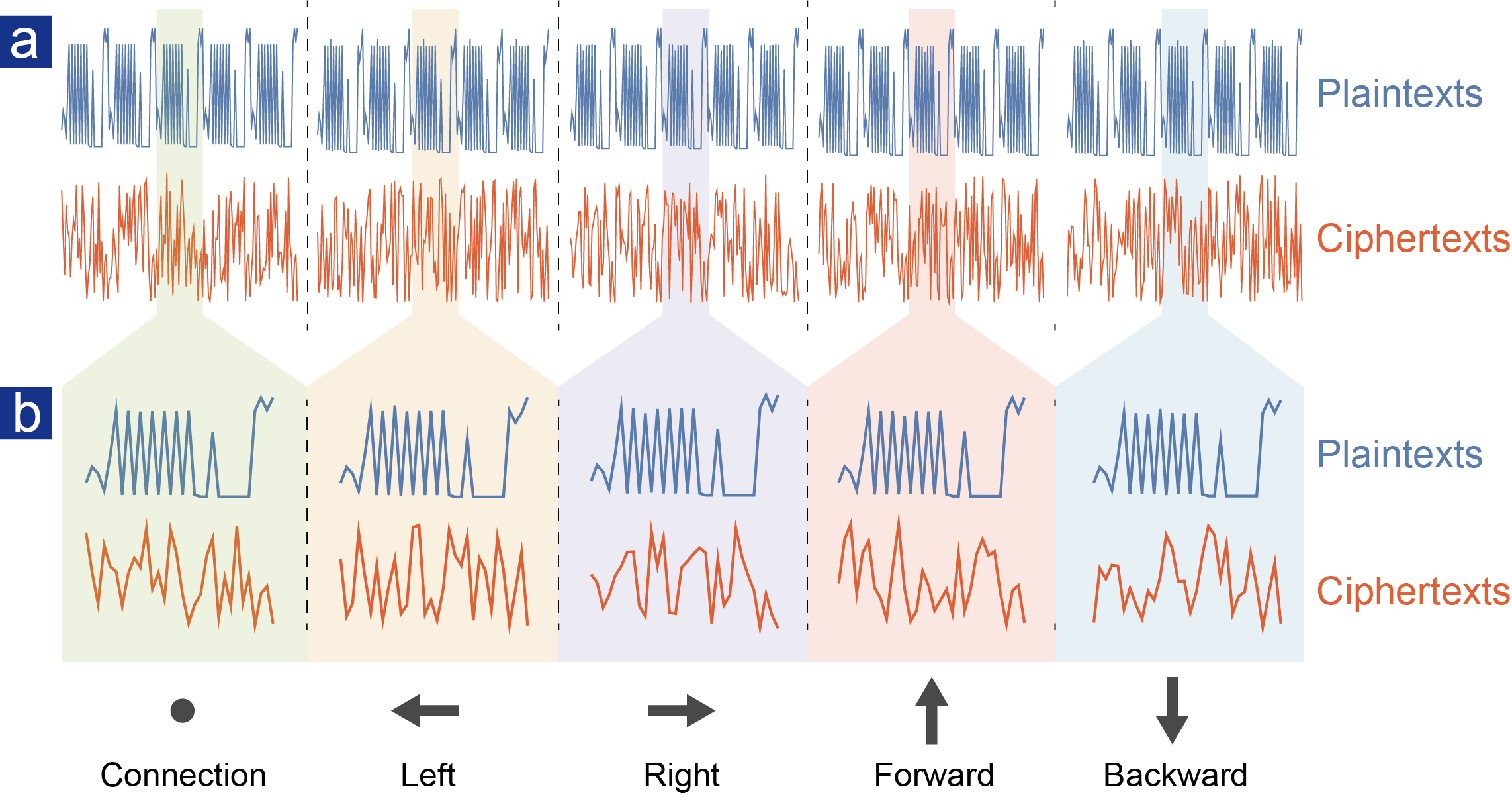}\\
	\caption{
		\textbf{Experimental results of intercepted commands} \textbf{(a)} Intercepted five pieces of typical commands with (blue lines) and without (red lines) one-time pad encryption for comparison. \textbf{(b)} Details of each functional command. The last four bytes of ciphertexts represent the assigned address information of quantum keys. One command is combined with 32 bytes, and each byte represents for an integer whose values range from 0 to 255, stored with eight binaries. See Supplementary Materials for details.}
	\label{f4}
\end{figure*}

The successful execution of one-time pad algorithm depends on the synchronization of keys. Unfortunately, it happens that commands get lost or error, leading to key mismatch between controllers and controlees. Any minor key mismatch may cause control system failure. To solve this problem, we assign unique address information to each command, so that each command with quantum keys is independent and well labeled. Once error happens, the corresponding command will be discarded directly together with its encryption and decryption keys to maintain the synchronization. Such address information doesn't have to be encrypted, because they include no information about commands.

The randomness of quantum keys is a crucial parameter that determines system security. Quantum random number based on the uncertainty principle of quantum mechanics provides the honest-to-goodness randomness in the world, with the properties of unpredictable and unreproducible \cite{Schneier_1996}. We use NIST suites to perform statistical tests \cite{Rukhin_APL_2015}, and the final results are shown in Figure 2(b). The results are $P\text{-}valuess$ of all 15 tests: indicating how a sequence is identical to purely random number, ideally $P\text{-}values$ equals to one. The results of NIST tests prove an excellent statistical randomness of our quantum keys.

Furthermore, according to one-time pad algorithm, where ciphertexts are the XOR values of quantum keys and plaintexts, the randomness of quantum keys determines that of ciphertexts. We intercept a section of commands in the air sent by the controller, and test them with three characteristics of random binary sequences proposed by Gobomb: balance property, runs property and auto-correlation property \cite{Golomb_SF_1967}. The good properties of balance, runs (see Table I), and auto-correlation (see Figure3) indicate that our ciphertexts are statistically random. Meanwhile, since the quantum keys are unpredictable and unrepeatable, the ciphertexts intercepted here are expected and experimentally observed to be truly random, which is impossible to be deciphered without matched quantum keys.

Experimental commands between the remote control and the aircraft are shown in Figure 4. We intercept five pieces of different functional commands, with and without one-time pad encryption for comparison. For plaintexts, we can see that five repeating commands share exactly identical values; while for ciphertexts, five repeating commands are bought into chaos, and there is no correlation between any two commands or even any two bytes, which guarantees the security as have been proved statistically in Table I and Figure 3, according to three postulates proposed by Golomb \cite{Golomb_SF_1967}.

Since the security of the commands depends on the one-time pad, the communication capacity in this cryptographic control scheme is mainly limited by the number of pre-charged QRNs. To extend the key updating period, on one hand, the capacity of secure key storage device should be large enough, while it might take more bytes in the commands for storing keys' address information. On the other hand, the encryption commands can be optimized according to different structures, and some trivial information in a certain command could be ignored to save keys, as well as to increase decryption speed.

In summary, we have proposed and experimentally demonstrated a quantum-enhanced cryptographic remote control scheme that combines quantum randomness and one-time pad algorithm for delivering commands remotely. The quantum-enhanced cryptographic scheme are expected to be generalized to bidirectional systems: controlees can be securely controlled and also be able to send encrypted recorded flight data back to controllers, such as position, direction and speed. Besides, the point-to-point solution can be extended to point-to-multipoint or distributed networks. More importantly, such scheme can be combined with fixed QKD channels \cite{Ursin_nphy_2007,Frohlich_PRA_2013,Marsili_nphoton_2013,Wang_nphoton_2013} for long-distance quantum keys charging, providing a flexible solution for control security of artificial intelligence and IoT in large scale. 

\section*{Acknowledgments}
The authors thank J.-W. Pan for helpful discussions. This work was supported by National Key R\&D Program of China (2017YFA0303700), the National Natural Science Foundation of China (NSFC) (61734005, 11761141014, 11690033), the Science and Technology Commission of Shanghai Municipality (STCSM) (15QA1402200, 16JC1400405, 17JC1400403), the Shanghai Municipal Education Commission (SMEC)(16SG09, 2017-01-07-00-02-E00049), and the Zhiyuan Scholar Program (ZIRC2016-01). X.-M.J. acknowledges support from the National Young 1000 Talents Plan.

\clearpage
\newpage

%%%%%%%%%%%%%%%%% Supplemental Information %%%%%%%%%%%%%%%%%%%%
\onecolumngrid
\subsection*{\large Supplemental Information: Experimental Quantum-enhanced Cryptographic Remote Control}
\setcounter{figure}{0}
\setcounter{table}{0}
\setcounter{equation}{0}
\renewcommand{\figurename}{Supplementary Figure}
\renewcommand{\tablename}{Supplementary Table}

\renewcommand{\thetable}{\arabic{table}}
\renewcommand{\theequation}{{S}\arabic{equation}}

\bigskip
\subsection*{\large Supplementary Note 1: Commands reliability testing}

To test the reliability of cryptographic remote control scheme, commands emitted by the UAV controller are intercepted and processed by another microcontroller (STM32). As for general cases, without encryption, the five-time repeated commands intercepted are demonstrated in FIG. S1(a). Inserted is the command structure which consists of 32 bytes, and each byte consists of eight ``0" or ``1" bits. Thus the value of a byte ranges from 0 to 255. The key address information added in the command is not encrypted, which contains no information without local quantum key storage devices. 

As for situations with quantum-enhanced encryption, commands sent by the UAV controller are encrypted by one-time pad algorithm. The ciphertexts intercepted are shown in FIG. S1(b). We can see that the five-time repeated commands has been brought into chaos, except for the bytes which carry key address information (marked in shadows). According to the key address information, the controllee can read out quantum keys and perform one-time pad decryption process. The randomness of ciphertexts has been tested and presented in Figure 3 and Table I in the main text.

\begin{figure*}[h]
	\centering
	\includegraphics[width=0.9\columnwidth]{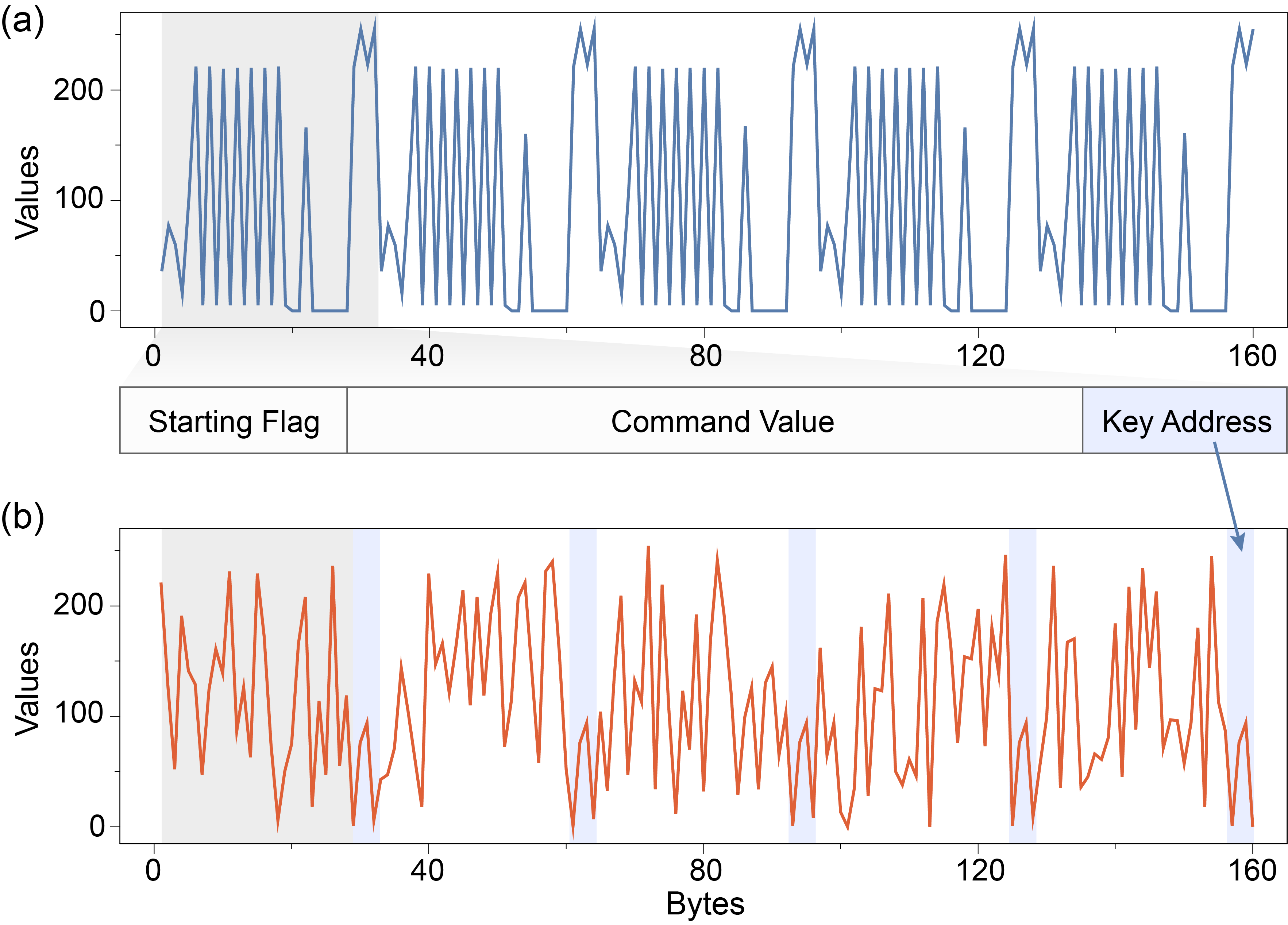}\\
	\caption{\textbf{Intercepted commands before and after encryption.} \textbf{(a)} The plaintexts of five-time repeated commands. The inserted represents the structure of a command. \textbf{(b)} The ciphertexts of five-time repeated commands. Shadows mark the key information in each commands.}
	\label{f1}
\end{figure*}

A brief structure of a command is shown in the insert of FIG. S1. Even different functional commands share parts of the same bytes to maintain stable communication process. Therefore, it is not necessary to encrypt all 32 bytes in a command, and we can just perform one-time pad encryption on the bytes that carry fatal information, to save quantum keys. Values of plaintexts of five commands (blue lines) presented in Figure 4 in the main text are shown in Table I. According to the command structure, they are similar but not the same.

\begin{table}
	\centering	
	\caption{\textbf{Intercepted values of plaintexts sent by the UAV controller. Red values show the difference.} }
	\label{tab1}
	\begin{tabular}{
			p{2.5cm}<{\centering}  p{0.8cm}<{\centering} 
			p{0.8cm}<{\centering}  p{0.8cm}<{\centering}
			p{0.8cm}<{\centering}  p{0.8cm}<{\centering}
			p{0.8cm}<{\centering}  p{0.8cm}<{\centering}
			p{0.8cm}<{\centering}  p{0.8cm}<{\centering}
			p{0.8cm}<{\centering}  p{0.8cm}<{\centering}
			p{0.8cm}<{\centering}  p{0.8cm}<{\centering}
			p{0.8cm}<{\centering}  p{0.8cm}<{\centering}
			p{0.8cm}<{\centering}  p{0.8cm}<{\centering}
			}

		\hline\noalign{\vskip 0.6mm}
		\hline\noalign{\vskip 0.14cm}
		
		\multirow{2}{*}{\textbf{Connection}} & 36 & 77 & 60 & 16 & 105 & \textcolor{red}{221} & 5 & \textcolor{red}{221} & 5 & \textcolor{red}{219} & \textcolor{red}{5} & \textcolor{red}{220} & \textcolor{red}{5} & \textcolor{red}{220} & 5 & \textcolor{red}{220}  \\
		 & 5 & \textcolor{red}{221}  & 5 & 0 & 0 & \textcolor{red}{166} & 0 & 0 & 0 & 0 & 0 & 0 & 221 & \textcolor{red}{255} & \textcolor{red}{223} & 255 \\

\hline\noalign{\vskip 0.6mm}
		
		\multirow{2}{*}{\textbf{Backward}} & 36 & 77 & 60 & 16 & 105 & \textcolor{red}{221} & 5 & \textcolor{red}{221} & 5 & \textcolor{red}{219} & \textcolor{red}{5} & \textcolor{red}{232} & \textcolor{red}{3} & \textcolor{red}{221} & 5 & \textcolor{red}{220}  \\
		 & 5 & \textcolor{red}{221}  & 5 & 0 & 0 & \textcolor{red}{149} & 0 & 0 & 0 & 0 & 0 & 0 & 221 & \textcolor{red}{191} & \textcolor{red}{215} & 255 \\

\hline\noalign{\vskip 0.6mm}
		
		\multirow{2}{*}{\textbf{Turn Left}} & 36 & 77 & 60 & 16 & 105 & \textcolor{red}{221} & 5 & \textcolor{red}{220} & 5 & \textcolor{red}{234} & \textcolor{red}{3} & \textcolor{red}{219} & \textcolor{red}{5} & \textcolor{red}{220} & 5 & \textcolor{red}{220}  \\
		 & 5 & \textcolor{red}{221}  & 5 & 0 & 0 & \textcolor{red}{168} & 0 & 0 & 0 & 0 & 0 & 0 & 221 & \textcolor{red}{255} & \textcolor{red}{215} & 255 \\

\hline\noalign{\vskip 0.6mm}
		
		\multirow{2}{*}{\textbf{Turn Right}} & 36 & 77 & 60 & 16 & 105 & \textcolor{red}{221} & 5 & \textcolor{red}{221} & 5 & \textcolor{red}{208} & \textcolor{red}{7} & \textcolor{red}{219} & \textcolor{red}{5} & \textcolor{red}{220} & 5 & \textcolor{red}{221}  \\
		 & 5 & \textcolor{red}{220}  & 5 & 0 & 0 & \textcolor{red}{168} & 0 & 0 & 0 & 0 & 0 & 0 & 221 & \textcolor{red}{255} & \textcolor{red}{215} & 255 \\

\hline\noalign{\vskip 0.6mm}
		
		\multirow{2}{*}{\textbf{Forward}} & 36 & 77 & 60 & 16 & 105 & \textcolor{red}{220} & 5 & \textcolor{red}{220} & 5 & \textcolor{red}{220} & \textcolor{red}{5} & \textcolor{red}{208} & \textcolor{red}{7} & \textcolor{red}{220} & 5 & \textcolor{red}{219}  \\
		 & 5 & \textcolor{red}{221}  & 5 & 0 & 0 & \textcolor{red}{168} & 0 & 0 & 0 & 0 & 0 & 0 & 221 & \textcolor{red}{255} & \textcolor{red}{215} & 255 \\
		
		\hline\noalign{\vskip 0.6mm}
		\hline\noalign{\vskip 0.14cm}		
	\end{tabular}
\end{table}


\begin{thebibliography}{99}

\bibitem{Schneier_1996} Schneier, B. \textit{Applied Cryptography, Second Edition: Protocols, Algorthms, and Source Code in C} (John Wiley \& Sons, Inc, 1996).

\bibitem{Gingerich_Isis_1996} Gingerich, O. The Codebreakers: The Codebreakers. The Story of Secret Writing, by David Kahn. \textit{Isis} (1996).

\bibitem{Shannon_Bell_1949} Shannon, C. E. Communication theory of secrecy systems. \textit{Bell Syst. Tech. J.} \textbf{28}, 656-715 (1949).

\bibitem{Knuth_1981} Knuth, D. E. \textit{The art of computer programming Volume 2: Seminumerical algorithms Reading} (And Searching, 1981).

% ----------- random number generator 
\bibitem{Jennewein_RSI_2000} Jennewein, T., Achleitner, U., Weihs, G., Weinfurter, H. \& Zeilinger, A. A fast and compact quantum random number generator. \textit{Rev. Sci. Instrum.} \textbf{71}, 1675-1680 (2000).

\bibitem{Stefanov_JMO_2000} Stefanov, A., Gisin, N., Guinnard, O., Guinnard, L. \& Zbinden, H. Optical quantum random number generator. \textit{J. Mod. Opt.} \textbf{47}, 595-598 (2000).

\bibitem{Dynes_APL_2008} Dynes, J. F., Yuan, Z. L., Sharpe, A. W. \& Shields, A. J. A high speed, postprocessing free, quantum random number generator. \textit{Appl. Phys. Lett.} \textbf{93}, 910 (2008).

\bibitem{Uchida_nphoton_2008} Uchida, A., Amano, K., Inoue, M., Hirano, K., Naito, S., Someya, H., Oowada, I., Kurashige, T., Shiki, M., Yoshimori, S., Yoshimura, K. \& Davis, P. Fast physical random bit generation with chaotic semiconductor lasers. \textit{Nature Photon.} \textbf{2}, 728-732 (2008).

\bibitem{Fiorentino_PRA_2006} Fiorentino, M., Santori, C., Spillane, S. M., Beausoleil, R. G. \& Munro, W. J. Secure self-calibrating quantum random-bit generator. \textit{Phys. Rev. A} \textbf{75}, 723-727 (2006).

% -------- physical RNG ----------

\bibitem{Stojanovski_IEEE_2001_partI} Stojanovski, T. \& Kocarev, L. Chaos-based random number generators-part I: analysis. \textit{IEEE Trans. Circ. Syst. I: Fund. Theory Appl.} \textbf{48}, 281-288 (2001).

\bibitem{Stojanovski_IEEE_2001_partII} Stojanovski, T., Pihl, J. \& Kocarev, L. Chaos-based random number generators-part II: practical realization. \textit{IEEE Trans. Circ. Syst. I: Fund. Theory Appl.} \textbf{48}, 382-385 (2001).

\bibitem{Petrie_Computers_2004} Petrie, C. S. \& Connelly, J. A. A noise-based IC random number generator for applications in cryptography. \textit{IEEE Trans. Circ. Syst. I: Fund. Theory Appl.} \textbf{47}, 615-621 (2000).

\bibitem{Szczepanski_IEEE_2000} Szczepanski, J., Wajnryb, E., Amig{\'o}, J. M., Sanchez-Vives, M. V. \& Slater, M. Biometric random number generators. \textit{Computers \& Security} \textbf{23}, 77-84 (2004).

\bibitem{Kohlbrenner_2004} Kohlbrenner, P. \& Gaj, K. An Embedded True
Random Number Generator for FPGAs. in \textit{Proc. of the 12th Int. Symp. on Field
Programmable Gate Arrays.} 71-78 (2004).

\bibitem{Herrero_RMP_2017} Herrero-Collantes, M. \& Garcia-Escartin, J. C. Quantum random number generators. \textit{Rev. Mod. Phys.} \textbf{89}, 015004 (2017).

% -------- QKD  ------------
\bibitem{Bennett_IEEE_1984} Bennett, C. H. \& Brassard, G. Quantum cryptography: public key distribution and coin tossing. \textit{Theoretical Computer Science} \textbf{560}, 7-11 (2014).

\bibitem{Ekert_PRL_1991} Ekert, A. K. Quantum cryptography based on Bell's theorem. \textit{Phys. Rev. Lett.} \textbf{67}, 661–663 (1991).

\bibitem{Lo_nphoton_2014} Lo, H.-K., Curty, M. \& Tamaki, K. Secure quantum key distribution. \textit{Nat. Photonics}. \textbf{8}, 595–604 (2014).

\bibitem{JiLing_OE_2016} Ji, L., Gao, J., Yang, A. L., Feng, Z., Lin, X. F., Li, H. G. \& Jin, X. M. Towards quantum communication in free-space seawater. \textit{Opt. Express} \textbf{25} (2016).

% -------- QKD networdk  -------------
\bibitem{Qiu_Nature_2014} Qiu \& Jane. Quantum communications leap out of the lab. \textit{Nature} \textbf{508}, 441-442 (2014).

\bibitem{Elliott_nCommun_2006} Elliott, C. The DARPA quantum network. \textit{Quantum Communications and cryptography} 83-102 (2006).

\bibitem{Fujiwara_OE_2011} Fujiwara, M., Ishizuka, H., Miki, S., Yamashita, T., Wang, Z., \& Tanaka, A.,\textit{ et al}. Field test of quantum key distribution in the Tokyo QKD network. \textit{Opt. Express } \textbf{19}, 10387-409 (2011).

\bibitem{Frohlich_PRA_2013} Fr\"{o}hlich, B., Dynes, J. F., Lucamarini, M., Sharpe, A. W., Yuan, Z., \& Shields, A. J.\textit{ et al.} A quantum access network. \textit{Nature} \textbf{501}, 69-72 (2013).

\bibitem{Tang_PRX_2016} Tang Y L, Yin, H. L., Zhao, Q., Liu, H., Sun, X. X., \& Huang, M. Q. Measurement-device-independent quantum key distribution over untrustful metropolitan network. \textit{Phys. Rev. X} \textbf{6}, 011024 (2016).

\bibitem{Gubbi_FGCY_2012} Gubbi, J., Buyya, R., Marusic, S., \& Palaniswami, M. Internet of Things (IoT): A vision, architectural elements, and future directions. \textit{Future Generation Computer Systems} \textbf{29}, 1645-1660 (2012).

\bibitem{Rukhin_APL_2015} Rukhin, A., Soto, J., Nechvatal, J., Smid, M., Barker, E., Leigh, S. \& Levenson, M. A statistical test suite for random and pseudorandom number generators for cryptographic applications. \textit{Appl. Phys. Lett.} \textbf{22}, 1645-179 (2015). 

\bibitem{Golomb_SF_1967} Golomb, S. W. \textit{Shift Register Sequences} (Holden-Day, 1967).

% ------------------------- Conclusion QKD channels --------------
\bibitem{Ursin_nphy_2007} Ursin, R. \textit{et al.} Entanglement-based quantum communication over 144km. \textit{Nature Phys.} \textbf{3}, 481-486 (2007).

\bibitem{Marsili_nphoton_2013} Marsili, F. \textit{et al.} Detecting single infrared photons with 93\% system efficiency. \textit{Nature Photon.} \textbf{7}, 210-214 (2013).

\bibitem{Wang_nphoton_2013} Wang, J.-Y. \textit{et al}. Direct and full-scale experimental verications towards ground-satellite quantum key distribution. \textit{Nature Photon}. \textbf{7}, 387-393 (2013).

\end{thebibliography}
\end{document}